\documentclass[twocolumn,nofootinbib,amsmath,prd,aps,superscriptaddress,tightenlines,preprintnumbers]{revtex4}
\usepackage{graphicx}
\usepackage{epstopdf}
\usepackage{amsfonts}
\usepackage{bm}
\usepackage{epsfig}
\usepackage{xspace}
\usepackage[usenames]{color}
\def\bea{\begin{eqnarray}}
\def\eea{\end{eqnarray}}
\def\ba{\begin{eqnarray}}
\def\ea{\end{eqnarray}}
\def\be{\begin{equation}}
\def\ee{\end{equation}}
\def\beq{\begin{equation}}
\def\eeq{\end{equation}}

\newcommand{\lsim}{\mathrel{\rlap{\lower4pt\hbox{\hskip1pt$\sim$}}
    \raise1pt\hbox{$<$}}}         
\newcommand{\gsim}{\mathrel{\rlap{\lower4pt\hbox{\hskip1pt$\sim$}}
    \raise1pt\hbox{$>$}}}         

\newcommand{\leftrightarrowraised}{\mathrel{\rlap{\lower-0pt\hbox{\hskip1pt$\partial$}}
    \raise6 pt\hbox{$\leftrightarrow$}}}

\begin{document}

\newcount\hour \newcount\minute
\hour=\time \divide \hour by 60
\minute=\time
\count99=\hour \multiply \count99 by -60 \advance \minute by \count99
\newcommand{\mydate}{\ \today \ - \number\hour :00}

\title{Testing for new physics in singly Cabibbo suppressed $D$ decays}

\def\Cincy{Department of Physics, University of Cincinnati, Cincinnati, Ohio 45221,USA}
\def\Cornell{Institute for High Energy Phenomenology, Newman Laboratory of 
Elementary Particle Physics, Cornell University, Ithaca, NY14853, USA}

\author{Yuval Grossman}
\email[Electronic address:]{yg73@cornell.edu} 
\affiliation{\Cornell}

\author{Alexander L. Kagan}
\email[Electronic address:]{kaganalexander@gmail.com} 
\affiliation{\Cincy}

\author{Jure Zupan}
\email[Electronic address:]{zupanje@ucmail.uc.edu}
\email[On leave of absence from Josef Stefan Institute and
U. of Ljubljana, Ljubljana, Slovenia.]{}
\affiliation{\Cincy}

\begin{abstract}
We devise tests for a new physics origin of the recently measured
direct CP violation in singly Cabibbo suppressed $D$ decays. The tests take the form of sum rules for the CP
asymmetries in various $D$ decays. They are based on the fact that
within the standard model CP violation arises from interference of the dominant tree amplitudes with the $\Delta I=1/2$ penguin amplitudes. The sum rules would be
violated if the observed CP violation is due to new physics
contributions to the effective weak Hamiltonian that change isospin by
$\Delta I=3/2$.
\end{abstract}

\maketitle
\newpage

\section{Introduction} 
Recently, the LHCb collaboration announced a measurement
of the difference between the time-integrated CP asymmetries in two singly Cabibbo
supressed (SCS) $D$ meson decay modes \cite{:2011in},
\beq
\begin{split}\label{DeltaACP1}
\Delta {\cal A}_{CP}&\equiv {\cal A}_{CP}(D\to K^+K^-) -
{\cal A}_{CP}(D\to \pi^+\pi^-) \\
&= (-0.82 \pm 0.21 \pm
0.11)\%,
\end{split}
\eeq
which has been confirmed by the CDF measurement~\cite{CDF-talk}
\beq \label{eq:CDFACP}
\Delta {\cal A}_{CP}= (-0.62 \pm 0.21 \pm
0.10)\%\,.
\eeq
The updated world average for the difference of the direct CP asymmetries is then $\Delta {\cal A}_{CP}^{\rm dir} =  (-0.67 \pm 0.16)\%$
\cite{CDF-talk}. In the SM, CP violation (CPV) in SCS $D$ decays comes from the
interference of the tree and penguin amplitudes and is
parametrically suppressed by $\mathcal O(V_{cb} V_{ub}/V_{cs}
V_{us})\sim 10^{-3}$.  However, the uncertainties on these order of magnitude estimates are large~\cite{Brod:2011re}.
In particular, the penguin contraction power corrections could be significantly
enhanced,  possibly leading to the observed CP asymmetry.

The predictions for direct CPV in charm decays are notoriously
difficult to make, even if one is aiming at order of magnitude
estimates. For instance, in \cite{Golden:1989qx} and recently in \cite{Brod:2011re, Pirtskhalava:2011va, Feldmann:2012js, Brod:2012ud} 
it was argued that large CP asymmetries can be expected in SCS $D$ decays.
In~\cite{Buccella:1994nf,Cheng:2012wr,Li:2012cf}, small CP asymmetries were obtained, while marginal consistency with measurements \eqref{DeltaACP1}, \eqref{eq:CDFACP} was found in \cite{Franco:2012ck}. At
the same time, there are viable new physics models (NP) that can enhance
$\Delta {\cal A}_{CP}$ and simultaneously avoid constraints from other
flavor violation searches, such as $D-\bar D$ mixing, as shown in
\cite{Feldmann:2012js,Grossman:2006jg,Isidori:2011qw,Wang:2011uu,Hochberg:2011ru,Chang:2012gn,Giudice:2012qq,Altmannshofer:2012ur,Chen:2012am,Hiller:2012wf}.

The aim of this work is to provide experimental tests that can
distinguish between a SM and NP origin for $\Delta {\cal A}_{CP}$.
The basic idea it to take advantage of the fact that the penguin
amplitudes are $\Delta I=1/2$ transitions. A prediction of
the SM is that CPV effects
are confined to the $\Delta I=1/2$
amplitude, to very good approximation (to be quantified below). Any observation of CPV effects due to the other possible isospin
reduced matrix element, namely the $\Delta I=3/2$ amplitude, would be a clear signal of new physics.

This insight can be used to experimentally search for NP in charm decays without relying on theoretical calculations. To do so we construct a set of CP asymmetry sum rules that will be obeyed if
the observed CP asymmetries are due to the SM, but violated if they
are due to NP. It is important to note that we can search in this way
only for a subset of NP models -- the ones that generate CPV in the $\Delta I=3/2$ reduced matrix elements. An example is a model with a single new
scalar field with nontrivial flavor couplings, as recently proposed in~\cite{Hochberg:2011ru} to explain the observed $\Delta {\mathcal A}_{CP}$.  On the other hand, NP models that only
contribute through penguin $\Delta I=1/2$ operators would not violate
the derived sum rules and are thus much harder to distinguish from the SM
contributions to $\Delta {\cal A}_{CP}$. Examples are provided by flavor violating supersymmetric squark-gluino loops that mediate the
$c\to u g$ transition~\cite{Grossman:2006jg,Giudice:2012qq,Hiller:2012wf}.

In the derivation of the sum rules we use the isospin flavor symmetry
of QCD. Isospin symmetry is broken at ${\mathcal O}(1\%)$, which is 
also the size of the CP asymmetries we are interested in. Thus special care is needed in order to avoid the introduction of
large errors in the sum rules. There are several 
sources of isospin breaking that could modify the sum rules. 
The $u$ and $d$ quark masses and electromagnetic interactions
break isospin in a CP conserving manner. Their effects are explicitly included in 
the sum rules so that they cancel up to quadratic order in isospin breaking.
The electroweak penguin operators provide CP violating sources for isospin breaking.
However, their effects are suppressed by $\alpha/\alpha_S\sim {\mathcal O}(10^{-2})$ compared to 
the leading CP violating but isospin conserving penguin contractions of the $Q_{1,2}$ operators, and can thus be safely neglected.

The paper is organized as follows. Our notation is introduced in Section \ref{Sec:Preliminaries}, 
the sum rules are derived in Section \ref{Sec:searching}, and we conclude in Section \ref{Conclusions}.

\section{Preliminaries}
\label{Sec:Preliminaries} 
The CP-conjugate decay amplitudes for SCS
decays can be written as 
\beq
\begin{split}
A_f   (D \to f )  &=   T_{f} +P_f e^{i(\delta^P_f-\gamma)}+A_f^{NP}e^{i(\delta^{NP}_f-\phi)},\\
\overline{A}_{\overline f}(\bar D \to \bar{f}  ) &=  T_{f}+P_f e^{i(\delta^P_f+\gamma)}+A_f^{NP}e^{i(\delta^{NP}_f+\phi)},
\end{split}
\eeq
where $T_{f} $ is the dominant SM ``tree" amplitude. It is proportional
to $V_{ud} V_{cd}^*$ and is taken to be real by convention.
The SM ``penguin" amplitude  has magnitude $P_f $. It is CKM suppressed by ${\mathcal O}(V_{cb} V_{ub}/V_{cs} V_{us})$
compared to the tree amplitude. It carries  the CKM weak phase
$\gamma=(67.3^{+4.2}_{-3.5})^\circ$~\cite{Charles:2011va} and the relative strong
phase $\delta_f $. 
The NP amplitude has magnitude $A_{f}^{NP}$, 
and carries a strong phase $\delta^{NP}$ and a weak phase $\phi$ relative
to the tree amplitude.

The direct CP asymmetry is given by
\beq\label{Adir.eq}
\begin{split}
{\cal A}_f^{\rm dir}   &\equiv \frac{|A_f |^2-| \bar A_{\bar f }|^2}
{|A_f |^2  + | \bar A_{\bar f } |^2}= \\
 & = 2 r_f^P \sin \gamma \sin \delta_f^P+2 r_f^{NP} \sin \phi \sin \delta_f^{NP},
 \end{split}
\eeq
where $r_f^P\equiv P_f/T_f$, $r_f^{NP}\equiv A_f^{NP}/T_f$, and we have
neglected higher orders in $r_f^{P}$ and $r_f^{NP}$. The question we
are interested in is how one can distinguish between the SM
contributions to the direct CP asymmetries, proportional to $r_f^P$, and
the NP contributions to the direct CP asymmetries, proportional to
$r_f^{NP}$. To do so, we will utilize the transformation properties of
the SM and NP contributions under isospin.

We first review the transformation properties of the SM contributions under isospin. The effective $\Delta C=1$ weak 
Hamiltonian, $H_{\rm eff}, $ is~\cite{Buchalla:1995vs}
\begin{equation}
\begin{split}
H_{\rm eff}^{\Delta C=1} &= \frac{G_F}{\sqrt{2}} \Big[\sum_{p=d,s} 
V_{cp}^* V_{up}  \left(C_1 Q_1^p + C_2 Q_2^p \right) \\
&-V_{cb}^* V_{ub} \sum_{i=3}^{6} 
C_i Q_i + C_{8g} Q_{8 g} \Big]
+ {\rm h.c.} .
\end{split}
\end{equation}
The ``tree" operators are 
\begin{equation}
\begin{split}
Q_1^p&=(\bar p c)_{V-A}(\bar u p)_{V-A}, \\
Q_2^p&=(\bar p_\alpha c_\beta)_{V-A} (\bar u_\beta p_\alpha)_{V-A}
\end{split}
\end{equation}
with summation over color indices $\alpha,\beta$ understood. The
QCD penguin operators are 
\begin{equation}
\begin{split}
Q_{3,5}&=(\bar u c)_{V-A} \sum_{q=u,d,s} (\bar qq)_{V\mp
  A},\\ Q_{4,6}&=(\bar u_\alpha c_\beta)_{V-A} \sum_{q=u,d,s} (\bar q_\beta
q_\alpha)_{V\mp A}, \\ 
Q_{8g} &= -\frac{g_s}{8\pi^2}\, m_c \bar u
\,\sigma_{\mu\nu}(1+\gamma_5) G^{\mu\nu} c\,,
\end{split}
\end{equation}
and we do not display the numerically further suppressed electroweak penguin operators.

The flavor structure of the tree operators for $D\to \pi\pi$ and $D_s\to
K\pi$ decays is $(\bar d c)(\bar u d)$.  Thus they have both $\Delta I=3/2$
and $\Delta I=1/2$ components. The remaining operators, i.e., 
the $(\bar s c)(\bar u s)$ tree operators for $D\to KK$ decays and the penguin operators, are purely
$\Delta I=1/2$.  Note that the penguin contraction contributions of the $(\bar d c)(\bar u d)$ tree operators to $D\to KK$ 
are $\Delta I =1/2$. 

The NP models that contribute to the CP asymmetries in SCS $D$ decays can
be grouped in to two sets, (i) those in which the NP operators are purely $\Delta
I=1/2$, and (ii) those in which the NP operators also have $\Delta I=3/2$ components.
As we show below, one can use the isospin decomposition and the resulting sum
rules to search for the presence of $\Delta I=3/2$ NP
just using experimental information.

\section{Searching for new physics via isospin}
\label{Sec:searching}
We now derive CP asymmetry sum rules that can be used to probe for the
presence of $\Delta I=3/2$ NP contributions.  Among the SCS decays,
the $D\to\pi\pi$, $D\to\rho\pi$, $D\to\rho\rho$, $D\to K\bar K\pi$ and $D_s\to K^{*}\pi$ modes
carry enough information for such tests. We discuss each
of them in turn.

\subsection{$D\to \pi\pi$ and $D\to \rho\rho$ decays}
The isospin decomposition of the $D^0\to \pi\pi$ decays is
\begin{subequations}\label{Apipi}
\begin{align}
A_{\pi^+\pi^-}&=\sqrt{2}{\cal A}_3+\sqrt2 {\cal A}_1,\\
A_{ \pi^0\pi^0}&=2 {\cal A}_3-{\cal A}_1,\\
A_{\pi^+\pi^0}&=3 {\cal A}_3,
\end{align}
\end{subequations} 
where ${\cal A}_3$ and ${\cal A}_1$ are the reduced matrix elements
for the $\Delta I=3/2$ and $\Delta I=1/2$ Hamiltonians. The phase convention used is such that $(u,d)$, $(\bar d, -\bar u)$, $(D^+,D^0)$,  $(K^+, K^0)$ and $(\bar K^0, K^-)$ form isospin doublets, while $(\pi^+,\pi^0,\pi^-)$ form a triplet. For the 
$\bar D^0\to
\pi\pi$ system the isospin decomposition is similar to \eqref{Apipi},
with ${\cal A}_3$, ${\cal A}_1$ replaced by the CP conjugate matrix elements
$\bar {\cal A}_3$, $\bar {\cal A}_1$.

We decompose the reduced matrix elements into SM
and NP contributions, with magnitudes $A_k$ and $a_k$,
respectively,
\beq \label{eq:not-sep}
{\cal A}_k= A_k e^{i(\delta_k^A-\phi_k^A)}+a_ke^{i(\delta_k^a-\phi_k^a)}, \qquad k=1,3.
\eeq
By convention we can set the strong phase $\delta_3^A=0$. In the SM
the weak phase of the $\Delta I=3/2$ amplitude is
also zero to excellent approximation, so that
we can set $\phi_3^A=0$.  Thus, in the SM the purely $\Delta I=3/2$ decay $D^+\to \pi^+\pi^0$ has
${\mathcal A}^{\rm dir}(D^+\to \pi^+\pi^0)=0$. However, the rate difference can be nonzero in the presence of NP, being given by
\beq \label{eq:dpipi}
|A_{\pi^+\pi^0}|^2-|\bar A_{\pi^-\pi^0}|^2= 
36 a_3  A_3 \sin\phi_3^a \sin \delta_3^a.
\eeq
Note that the CP asymmetry is proportional to the $\Delta I=3/2$ NP
coefficient $a_3$. 

Let us comment on the isospin
breaking effects that have been ignored in the decomposition of \eqref{Apipi}. The isospin breaking due to the $u$, $d$ quark masses and due to the electromagnetic interactions can be safely neglected since they are CP
conserving.  Thus, they only modify ${\mathcal A}_{CP}(D^+\to \pi^+\pi^0)$ at
second order in small parameters. While ${\mathcal A}_{CP}(D^+\to
\pi^+\pi^0)\sim {\mathcal O}(r_f^{NP})$, the effect of isospin
breaking is ${\mathcal O}(\epsilon_I r_f^{NP, EWP} )$, where
$\epsilon_I$ is the typical size of isospin breaking. It is of order
$1\%$ and may be enhanced by at most a factor of a few. Similarly the
electroweak penguins can be neglected due to the small sizes of their
Wilson coefficients. Thus, we conclude that a measured nonzero CP
asymmetry in $D^+ \to \pi^+\pi^0$ would be a signal for $\Delta I=3/2$
NP.

Note that if a direct CP asymmetry is not found in $D^+\to \pi^+\pi^0$, this does
not mean that $\Delta {\mathcal A}_{CP}$ cannot be due to a new $\Delta I=3/2$ amplitude. It is
possible, for instance, that the strong phase difference $\delta_3^a$
between the NP and SM $\Delta I=3/2$ amplitudes is simply smaller
than the strong phase difference between the $\Delta I=3/2$ and $\Delta I=1/2$ amplitudes. 

We therefore devise two more tests for the presence of new CP violating
phases in the $\Delta I=3/2$ operators. The first involves the sum of
rate differences
\beq\label{sum:rate:diff}
\begin{split}
|&A_{\pi^+\pi^-}|^2-|\bar A_{\pi^-\pi^+}|^2+|A_{\pi^0\pi^0}|^2-|\bar
A_{\pi^0\pi^0}|^2\\
 &-\frac{2}{3}\big(|A_{\pi^+\pi^0}|^2-|\bar A_{\pi^-\pi^0}|^2\big) = 3 \left(|{\cal A}_1|^2- |\bar {\cal A}_1|^2\right).
 \end{split}
\eeq
The important point is that this sum only depends on the $\Delta I=1/2$
amplitudes.  Thus, if the sum is found to be nonzero this means that
there are $\Delta I=1/2$ contributions to the CP asymmetries. They
could be due to NP or they could be due to the SM. However, if the sum \eqref{sum:rate:diff} is found to
be zero, while the individual rate differences are nonzero, this would indicate
that the CP asymmetries are likely dominated by $\Delta I=3/2$ NP
contributions. This statement does come with a caveat. It would still be possible that, whereas the CPV weak phases are only present in the $\Delta I=1/2$ amplitude,
the strong phases between terms in ${\cal A}_1$ with different weak phases are small. In this case, ${\cal A}_{CP}(\pi^+\pi^-)$, and ${\cal A}_{CP}(\pi^0\pi^0)$ would be nonzero due to interference of the $\Delta I=1/2$ and $\Delta I=3/2$ amplitudes.

This possibility can be checked with more data if time dependent
$D(t)\to \pi^+\pi^-$ and $D(t)\to \pi^0\pi^0$ measurements become
available, or if there is additional information on relative phases
from a charm factory running on the $\Psi(3770)$ (for feasibility see, e.g. \cite{Bevan:2011up}). It amounts to measuring
the weak phase of the $\Delta I=3/2$ amplitude ${\cal A}_3$ via 
generalized triangle constructions that also take isospin
breaking into account. From the isospin decomposition we have an isospin
sum rule
\beq
\begin{split}
\frac{1}{\sqrt2} &A_{\pi^+\pi^-}+A_{\pi^0\pi^0}-A_{\pi^+\pi^0}=A_{\rm break},
\end{split}
\eeq
and a similar sum rule for the CP-conjugate decays. The amplitude $A_{\rm break}$ is due to isospin breaking and is of order ${\mathcal O}(\epsilon_I A_{i})$. 
It is equal in $D\to \pi\pi$ and $\bar D\to \pi\pi$ decays,  i.e., $A_{\rm break}=\bar A_{\rm break}$, up to very small CP violating corrections which are down by an extra factor of $r_f\lesssim {\cal O}(0.01)$.  One therefore has the following sum rule, valid even in the presence of isospin breaking,
\beq\label{triangle-rel}
\begin{split}
&\frac{1}{\sqrt2} A_{\pi^+\pi^-}+A_{\pi^0\pi^0} -\frac{1}{\sqrt2} \bar A_{\pi^-\pi^+}-\bar A_{\pi^0\pi^0}=\\
&~~3\big({\cal A}_3-\bar{\cal A}_3\big)=-6i a_3
e^{i\delta^a_3} \sin\phi^a_3,
\end{split}
\eeq
where in the last stage we use the fact that $A_3$ carries a negligible CP violating phase
in the SM.  Note that isospin breaking in this relation has canceled (up to corrections quadratic in small parameters). Therefore, if 
\beq\label{triangle-rel2}
\begin{split}
\frac{1}{\sqrt2} \big(A_{\pi^+\pi^-}- \bar A_{\pi^-\pi^+}\big)\ne -\big(A_{\pi^0\pi^0} -\bar A_{\pi^0\pi^0}\big),
\end{split}
\eeq
is found, this would mean there is CPV NP  in the $\Delta I=3/2$ amplitude. The relative phases between the $A_{\pi^+\pi^-}$ and  $\bar A_{\pi^-\pi^+}$ amplitudes and between the $ A_{\pi^0\pi^0}$ and $\bar A_{\pi^0\pi^0}$ amplitudes can be measured in entangled $\psi(3770)\to D\bar D$ decays. 
In addition, the phase between the $A_{\pi^+\pi^-}$ and
$\bar{A}_{\pi^-\pi^+}$ amplitudes can be obtained from the time
dependent $D(t)\to \pi^+\pi^-$ decay.   Similarly, the phase between
the $ A_{\pi^0\pi^0}$ and $\bar A_{\pi^0\pi^0}$ amplitudes can be
obtained from the time dependent $D(t)\to \pi^0\pi^0$ decay. The magnitudes of the amplitudes can be measured in their respective time integrated decays. We can thus form an experimental test. If
\beq\label{triangle-rel3}
\begin{split}
\frac{1}{\sqrt2} \big|A_{\pi^+\pi^-}- \bar A_{\pi^-\pi^+}\big|\ne \big|A_{\pi^0\pi^0} -\bar A_{\pi^0\pi^0}\big|,
\end{split}
\eeq
then a $\Delta I=3/2$ NP amplitude has been discovered.

While the above formalism has been written down for $D\to \pi\pi$ decays, it applies without changes to $D\to \rho\rho$ decays, but for each polarization amplitude separately. As long as the polarizations of the $\rho$ resonances are measured (or if the longitudinal decay modes dominate, as is the case in $B\to \rho\rho$ decays), the search for $\Delta I=3/2$ NP could be easier experimentally in $D\to \rho \rho$ decays. 

\subsection{$D\to \rho\pi$ decays}
Another experimentally favorable probe is the isospin analysis of the $D\to \pi^+\pi^-\pi^0$ Dalitz plot in terms of the $D\to \rho\pi$ decays. 
The isospin decomposition for $D^0$ decays is 
\begin{align}
A_{\rho^+\pi^-}&={\cal A}_3+{\cal B}_3+\frac{1}{\sqrt2}{\cal A}_1+{\cal B}_1,\\
A_{\rho^0\pi^0}&=2{\cal A}_3-{\cal B}_1,\\
A_{\rho^-\pi^+}&={\cal A}_3-{\cal B}_3-\frac{1}{\sqrt2}{\cal A}_1+{\cal B}_1,
\end{align}
and for $D^+$ decays it is
\begin{align}
A_{\rho^+\pi^0}&={\frac{3}{\sqrt2}}{\cal A}_3-\frac{1}{\sqrt2}{\cal B}_3+{\cal A}_1,\\
A_{\rho^0\pi^+}&={\frac{3}{\sqrt2}}{\cal A}_3+\frac{1}{\sqrt2}{\cal B}_3-{\cal A}_1,
\end{align}
where ${\cal A}_{3}, {\cal B}_3$ are the $\Delta I=3/2$ amplitudes for $I=2,1$ final states, while ${\cal A}_1, {\cal B}_1$ are $\Delta I=1/2$ amplitudes for $I=1,0$ final states, and we have assumed that the $\Delta I=5/2$ amplitude is negligibly small (since these interactions are small and CP conserving they would introduce corrections to our result that are only quadratic in small parameters).
The $D\to \pi^+\pi^-\pi^0$ decay is dominated by the isospin 0 final state \cite{Gaspero:2008rs}, which means that the reduced amplitude ${\cal B}_1$ is expected to be the largest. 

From the Dalitz plot for $D^0\to \pi^+\pi^-\pi^0$ one can measure the
relative phases of $A_{\rho^+\pi^-}, A_{\rho^0\pi^0}$ and
$A_{\rho^-\pi^+}$, as well as their magnitudes. The sensitivity to
phases comes from the overlaps of the $\rho$ resonances in the Dalitz
plot. This means that the magnitudes and phases (up to an overall
phase) of the reduced matrix elements ${\cal B}_1$, ${\cal A}_3$, and
${\cal B}_3+{\cal A}_1/\sqrt2$ are measurable. If the time dependent
Dalitz plot is measured then the relative phases between the ${\cal
A}_i$ and CP conjugate $\bar {\cal A}_i$ could be measured.

We first discuss CP asymmetry sum rules that can be obtained from time
integrated Dalitz plot measurements.  We again employ a notation in which 
the strong and weak phases of the SM and NP contributions appear explicitly, as in \eqref{eq:not-sep}.  The notation we use is the straightforward generalization of  \eqref{eq:not-sep}, with $\delta^{A}_{1,3}$, $\delta^{B}_{1,3}$ the SM strong phases, $\phi^{A}_{1,3}$, $\phi^{B}_{1,3}$ the  SM weak phases, $A_{1,3}$, $B_{1,3}$ the magnitudes of the SM reduced amplitudes, while NP contributions are denoted by small letters, $A\to a$, $B\to b$. By convention, the strong phase of the
SM amplitude $A_3$ is taken to be zero, $\delta_3^A=0$.  The weak phases
of the SM tree amplitudes $A_3$ and $B_3$ are also zero, $\phi_3^A=\phi_3^B=0$. There are two
combinations of measured amplitudes that are proportional to $\Delta
I=3/2$ amplitudes
\beq\begin{split}
A_{\rho^+\pi^0}+A_{\rho^0\pi^+}&=3 \sqrt2 {\cal A}_3,\\
A_{\rho^+\pi^-}+2A_{\rho^0\pi^0}+A_{\rho^-\pi^+}&=6{\cal A}_3.
\end{split}\eeq 
A measurement of the second sum can be obtained from the $D^0\to \pi^+\pi^-\pi^0$ Dalitz plot.
If the related CP asymmetry 
\beq
\begin{split}
|A_{\rho^+\pi^-}+&2A_{\rho^0\pi^0}\!+\!A_{\rho^-\pi^+}\!|^2-
|\overline{A}_{\rho^-\pi^+}+2\overline{A}_{\rho^0\pi^0}\!+\!\overline{A}_{\rho^+\pi^-}\!|^2
\\
=&36\big(|{\cal A}_3|^2-|\bar {\cal A}_3|^2\big)=144 A_3 a_3\sin\phi_3^a \sin\delta_3^a,\label{Arhopi-sumrule}
\end{split}
\eeq
is found to be nonzero, this would mean that the NP contribution $a_3$ is nonzero. If it is found to vanish, it could still be that this is due to the strong phase difference $\delta_3^a$ being vanishingly small. 

Assuming that this is the case, i.e. that $\delta_3^a=0$, one can still test for the presence of $\Delta I=3/2$ CP violating NP. The weighted sum 
\beq
\begin{split}\label{eq:sumrhopi}
&2\big(|A_{\rho^0\pi^0}|^2-|\bar A_{\rho^0\pi^0}|^2\big)+\\
&|A_{\rho^+\pi^-}+A_{\rho^-\pi^+}|^2-|\bar A_{\rho^+\pi^-}+\bar A_{\rho^-\pi^+}|^2\\
&=12 \big(|{\cal A}_3|^2-|\bar {\cal A}_3|^2\big)+6\big(|{\cal B}_1|^2-|\bar {\cal B}_1|^2\big),
\end{split}
\eeq
measures whether there is direct CP violation in the ${\cal A}_3$ or ${\cal B}_1$ reduced amplitudes. Let us assume that \eqref{Arhopi-sumrule} 
is found to be vanishingly small, so that $|{\cal A}_3|=|\bar {\cal A}_3|$. If the sum \eqref{eq:sumrhopi} is found to be zero as well, while the individual CP asymmetries are nonzero, this would be a strong indication for $\Delta I=3/2$ NP.  Again,as in the case of $\pi\pi$, there is a caveat, namely that it is possible that there is no direct CPV in ${\cal B}_1$ even though there are weak phases in ${\cal B}_1$. For instance, this would be the case if the strong phases for terms with different weak phases in ${\cal B}_1$ would be the same.  The individual CP asymmetries would then be nonzero due to interference of ${\cal B}_1$ with the other amplitudes, rather than $\Delta I=3/2$ NP.

A definitive answer can be provided by another test that is directly sensitive to the weak phase of ${\cal A}_3$. This test is possible if the time dependent $D(t)\to \pi^+\pi^-\pi^0$ Dalitz plot is measured. In this case the relative phases between the $D^0\to \rho\pi$ and $\bar D^0\to \rho\pi$ amplitudes can be obtained (alternatively one could use time integrated entangled decays of $\psi(3770)$ at the charm factory). 
The presence of a weak phase in ${\cal A}_3$ can then be determined from the following sum-rule
\beq\label{a3rhopi}
\begin{split}
&\big(A_{\rho^+\pi^-}+A_{\rho^-\pi^+}+2 A_{\rho^0\pi^0}\big)-\\
&\big(\bar A_{\rho^-\pi^+}+\bar A_{\rho^+\pi^-}+2 \bar A_{\rho^0\pi^0}\big)=\\
&~~~~~~~~6 \big({\cal A}_3-\bar{\cal A}_3\big)=-12i a_3
e^{i\delta^a_3} \sin\phi^a_3,
\end{split}
\eeq
where in the last stage we use the fact that the SM amplitude $A_3$ does not carry a
weak phase.  Thus, a non-vanishing result for \eqref{a3rhopi} would provide definitive proof for $\Delta I=3/2$ NP. A similar sum rule for the CP asymmetries rather than the amplitudes
was given in \eqref{Arhopi-sumrule}. In that case the time
integrated Dalitz plot suffices to determine the sum
rule inputs.

\subsection{$D \to K \bar K \pi$ decays}
The isospin decomposition for the $D^0$ decays is 
\begin{align}
A_{K^+ {\bar K^0} \pi^-}&={\cal B}_1-{\cal A}_1 +{\cal C}_3 +{\cal B}_3,\\
A_{K^+K^-\pi^0}&={\cal B}_1'+\frac{1}{\sqrt2}{\cal A}_1 +\sqrt2{\cal C}_3 +{\cal B}_3',\\ 
A_{K^0 {\bar K^0}\pi^0}&=-{\cal B}_1'+\frac{1}{\sqrt2}{\cal A}_1 +\sqrt2{\cal C}_3 -{\cal B}_3',\\ 
 A_{K^0 K^-\pi^+}&=-{\cal B}_1-{\cal A}_1 +{\cal C}_3 -{\cal B}_3,
\end{align}
and for the $D^+$ decays it is
\begin{align}
A_{K^+{\bar K^0} \pi^0}&=\sqrt2 {\cal B}_1 +\frac{3}{\sqrt2}{\cal C}_3 -\frac{1}{\sqrt2}{\cal B}_3,\\
A_{K^+ K^-\pi^+}&=-{\cal B}_1+\sqrt 2 {\cal B}_1'+\frac{3}{2}{\cal C}_3 +\frac12 {\cal B}_3-\frac{1}{\sqrt2}{\cal B}_3',\\ 
A_{K^0 \bar K^0\pi^+}&=-{\cal B}_1-\sqrt 2 {\cal B}_1'+\frac{3}{2}{\cal C}_3 +\frac12 {\cal B}_3+\frac{1}{\sqrt2}{\cal B}_3',
\end{align}
where ${\cal B}_3, {\cal B}'_3$, ${\cal C}_3$ are $\Delta I=3/2$ amplitudes for $I=1,1,2$ final states with the two kaons in the $I=1,0,1$ isospin state, while  ${\cal A}_1, {\cal B}_1$, ${\cal B}'_1$ are $\Delta I=1/2$ amplitudes for $I=0,1,1$ final states with the two kaons in $I=1,1,0$ isospin state.
The same results apply for $D\to K^*\bar K\pi$, $D\to K\bar K^*\pi$ and $D\to K^*\bar K^*\pi$ decays (and decays with a $\rho$ instead of a $\pi$ in the final state) with obvious replacements. 

In the case of $D^+$ decays it is possible to construct a purely $\Delta I=3/2$ matrix element by summing only three decay amplitudes, while in the case of $D^0$ decays four amplitudes are needed. For this reason we only consider the $D^+$ decays. For instance, for $D^+$ decays to $K^*\bar K^*$ resonances we have for each polarization (to shorten the notation we do not show the polarizations explicitly)
\beq
\begin{split}
\sqrt2 &A_{K^{*+}{\bar K^{*0}} \pi^0}+A_{K^{*+} K^{*-}\pi^+}+A_{K^{*0} \bar K^{*0}\pi^+}= 6 \,{\cal C}_3.
\end{split}
\eeq
Thus, if the CP violating difference
\beq\label{testK*K*}
\begin{split}
&|\sqrt2 A_{K^{*+}{\bar K^{*0}} \pi^0}+A_{K^{*+} K^{*-}\pi^+}+A_{K^{*0} \bar K^{*0}\pi^+}|^2 \\
&-|\sqrt2 \bar A_{K^{*-}{ K^{*0}} \pi^0}+\bar A_{K^{*-} K^{*+}\pi^-}+\bar A_{\bar K^{*0} K^{*0}\pi^-}|^2,
\end{split}
\eeq
is found to be nonzero, this would mean that there is $\Delta I=3/2$ NP. The relative phases of the three amplitudes can be measured in the five-body decay $D^+\to K^0 K^- \pi^0\pi^+\pi^+$ and its CP conjugate. All three resonant decays, $D^+\to {K^{*+}{\bar K^{*0}} \pi^0}, D^+\to {K^{*+} K^{*-}\pi^+}$ and $D^+\to {K^{*0} \bar K^{*0}\pi^+}$ are part of this final state. The relative phases between the amplitudes can then be obtained from the overlaps of the resonances in the five body final state phase space. 

A somewhat more complicated possibility is represented by the $D\to K \bar K^*\pi$ and $D\to K^* \bar K \pi$ decays. A test that is similar to \eqref{testK*K*} can be devised for each of the two sets of decays. If either one of the CP violating differences 
\beq
\begin{split}\label{eq1sum}
&|\sqrt2 A_{K^{+}{\bar K^{*0}} \pi^0}+A_{K^{+} K^{*-}\pi^+}+A_{K^{0} \bar K^{*0}\pi^+}|^2 \\
&-|\sqrt2 \bar A_{K^{-}{ K^{*0}} \pi^0}+\bar A_{K^{-} K^{*+}\pi^-}+\bar A_{\bar K^{0} K^{*0}\pi^-}|^2,
\end{split}
\eeq
and 
\beq
\begin{split}\label{eq2sum}
&|\sqrt2 A_{K^{*+}{\bar K^{0}} \pi^0}+A_{K^{*+} K^{-}\pi^+}+A_{K^{*0} \bar K^{0}\pi^+}|^2 \\
&-|\sqrt2 \bar A_{K^{*-}{ K^{0}} \pi^0}+\bar A_{K^{*-} K^{+}\pi^-}+\bar A_{K^{*-} K^{+}\pi^-}|^2,
\end{split}
\eeq
is found to be nonzero, this would mean that there is $\Delta I=3/2$ NP. 

In order to experimentally construct  \eqref{eq1sum} or \eqref{eq2sum}, the magnitudes of the amplitudes and their relative phases need to be measured. To determine the relative phase differences a number of four body decays and their CP conjugates need to be measured. The phase difference between $A_{ K^{*+}K^-\pi^+}$  and $A_{ K^0\bar K^{0*} \pi^+}$  can be measured from the decay $D^+\to K^0K^- \pi^+\pi^+$ (the two amplitudes appear in \eqref{eq2sum} and \eqref{eq1sum}, respectively). The phase difference between $A_{K^{+}K^{*-}\pi^+}$ and $A_{K^{0*}\bar K^{0} \pi^+}$ can be measured from the decay $D^+\to K^+\bar K^0 \pi^-\pi^+$ (they appear in \eqref{eq1sum} and \eqref{eq2sum}, respectively). In order to completely fix all of the required phase differences, the decay $D^+\to K^0\bar K^0\pi^0\pi^+$ or the decay $D^+\to K^+ K^-\pi^0\pi^+$ also needs to be measured (as well as the CP conjugated decays of all the above mentioned modes). From the resonance overlaps in the decay $D^+\to K^0\bar K^0\pi^0\pi^+$, the relative phases of $A_{K^{*0}\bar K^{0} \pi^+}$, $A_{K^{*+}\bar K^{0} \pi^0}$ and $A_{K^{0}\bar K^{*0} \pi^+}$ can be obtained, so that  \eqref{eq2sum} is fully determined. Similarly, from the decay $D^+\to K^+ K^-\pi^0\pi^+$ the relative phases of $A_{K^{+}\bar K^{*0} \pi^0}$, $A_{K^{*+} K^{-} \pi^+}$ and $A_{K^{+}K^{*-} \pi^+}$ can be obtained so that, \eqref{eq1sum} is fully determined.

\subsection{$D_s$ decays}
It is also possible to search for CP violation in $\Delta I=3/2$ amplitudes using $D_s^+\to K^*\pi$ decays. The isospin decomposition is
\beq
\begin{split}
A(D_s^+\to \pi^0 K^{*+})&=\sqrt2 {\cal A}_3 -{\cal A}_1,\\
A(D_s^+\to \pi^+ K^{*0})&={\cal A}_3 +\sqrt2 {\cal A}_1.
\end{split}
\eeq
The two decays can be measured from the common Dalitz plot for $D_s^+\to K_S\pi^+\pi^0$, which has $K^{*+}$ and $K^{*0}$ bands 
that overlap with the $\rho^+$ band, while the two $K^*$ bands do no overlap directly. From the Dalitz plot analysis one can deduce the phase difference between the two amplitudes and construct the quantity
\beq
\sqrt2 A(D_s^+\to \pi^0 K^{*+})+A(D_s^+\to \pi^+ K^{*0})=3 {\cal A}_3.
\eeq
Direct CP violation in this sum, i.e., 
\beq
\begin{split}\label{DCPVDs}
&|\sqrt2 A(D_s^+\to \pi^0 K^{*+})+A(D_s^+\to \pi^+ K^{*0})|^2-\\
&|\sqrt2 A(D_s^-\to \pi^0 K^{*-})+A(D_s^-\to \pi^-{ \bar K^{*0}})|^2\ne 0,
\end{split}
\eeq
would necessarily be due to $\Delta I=3/2$ NP contributions. Additional information on the absolute value of $|A(D_s^+\to \pi^+ K^{*0})|$ can be obtained from the $D_s^+\to \pi^+ K^+\pi^-$ three body decay. 

An analogous test using $D_s \to \rho K^*$ decays also exists, with expressions obtained from the above via the replacement $\pi\to \rho$ and valid for each polarization separately. 
The relative phase between $A(D_s^+\to \rho^0 K^{*+})$ and $A(D_s^+\to \rho^+ K^{*0})$ can be measured from the four body decay $D_s^+\to \pi^+\pi^-K^+\pi^0$. 
The absolute magnitude $|A(D_s^+\to \rho^0 K^{*+})|$  can be obtained from the more easily measured decay $D_s^+\to \pi^+\pi^- K_S \pi^+$, and can be used as a further constraint. 

In most of the manuscript we kept the  final states $K^0$ and $\bar K^0$ mesons explicit in the notation. When measurements are  performed they will be part of the $K_S$ meson.
In checking for the presence of  $\Delta I=3/2$  NP one thus needs to keep track of the CP violation in the neutral kaon system.  This effect cannot be neglected as it generates CP
asymmetries of order $\epsilon_K$. However, this effect can be taken into
account explicitly by appropriately modifying the above sum rule equations and also by correcting for the time dependence efficiency for detecting the $K_S$~\cite{Grossman:2011zk}.

\section{Conclusions}
\label{Conclusions}
We have presented a set of isospin sum rules for CP asymmetries in singly Cabibbo suppressed $D$ decays that can be used to test for NP explanations of the measured  $\Delta {\cal A}_{CP}={\cal A}_{CP}(D\to K^+K^-) - {\cal A}_{CP}(D\to \pi^+\pi^-) $ that originate from a $\Delta I=3/2$ matrix element. The simplest test only requires the measurement of ${\cal A}_{CP}(D^+\to \pi^+\pi^0)$. If this is found to be nonzero then one has discovered NP in the $\Delta I=3/2$ transition. The same is true if  ${\cal A}_{CP}(D^+\to \rho^+\rho^0)\ne 0$ is found. Similar sum rules involving several $D\to\pi\pi, \rho \rho, \rho \pi, K^{(*)}\bar K^{(*)}\pi,  K^{(*)}\bar K^{(*)}\rho$ or $D_s\to K^*\pi,K^*\rho$ decay amplitudes were also derived. The isospin sum rules \eqref{Arhopi-sumrule}, \eqref{eq1sum}, \eqref{eq2sum}, \eqref{DCPVDs} only require time integrated measurements, while the isospin sum rules \eqref{triangle-rel3},  \eqref{a3rhopi} need time dependent measurements. Generically, if this type of NP is responsible for the bulk of the measured $\Delta {\cal A}_{CP}$, then violations of the isospin sum rules at the order of $\sim {\mathcal O}(0.5\%)$ can be expected, while the sum rules would be zero in the SM, up to corrections that are second order in isospin breaking.

\section*{Acknowledgements}
We thank Brian Meadows and Michael D. Sokoloff for useful discussions. Y.~G. is supported in part by the NSF grant PHY-0757868 and by a grant
from the BSF. A.~K. is supported by DOE grant FG02-84-ER40153. This work was facilitated in part by the workshop "New Physics from Heavy Quarks in Hadron Colliders"
which was  sponsored by the University of Washington  and supported
by the DOE under contract  DE-FG02-96ER40956.


\end{document}